\documentclass[seceqn]{elsart5p}

\usepackage{natbib}
\usepackage{graphicx}
\usepackage{amssymb}
\usepackage{lineno}

\newcommand{\pderiv}[2]{\frac{\partial #1}{\partial #2}}

\begin{document}

\begin{frontmatter}
 \title{Allee Effects and Extinction in a Lattice Model}
 \author{Alastair Windus and Henrik Jeldtoft Jensen\corauthref{cor1}}
 \corauth[cor1]{Corresponding author.
 Tel: +44 20 7594 8541, fax: +44 20 6594 8516, email address: h.jensen@imperial.ac.uk,
 h.jensen@ic.ac.uk (HJ. Jensen).}
 \address{Mathematics Department, Imperial College London, South Kensington  Campus, London. SW7 2AZ.}

\begin{abstract}
In the interest of conservation, the importance of having a large habitat available for a species is widely known. Here, we introduce a lattice-based model for
a population and look at the importance of fluctuations as well as that of the population density, particularly
with respect to Allee effects. We examine
the model analytically and by Monte Carlo simulations and find
that, while the size of the habitat is important, there exists a critical population density below which extinction is
assured. This has large consequences with respect to conservation, especially
in the design of habitats and for populations whose density has
become small. In particular, we find that the probability of survival for
small populations can be increased by a reduction in the size of the habitat and show that there exists an optimal size reduction.
\end{abstract}


\end{frontmatter}

\section{Introduction} \label{Section: Introduction}
Extinction is becoming a greater and greater issue all over
the world and is a cause of extreme concern. It has been estimated
that anthropogenic extinctions are resulting in the loss of a few percent of the current world's biosphere, which is of magnitude 3 to 4 times the natural background rate \citep{May}. 
The World Conservation Union
(IUCN), through its Species Survival Commission (SCC) develops criteria to assess the extinction rate for plants and animals
all over the world which enables them to keep a so-called \textit{Red List} (www.redlist.org) of species which are threatened with extinction in order to promote their
conservation. The list currently shows over 16,000 threatened species around the world - a 45\% increase on the figure from the year 2000. 

It has been shown by examining both  discrete \citep{Escudero} and continuous
\citep{SKellam} populations that, at least analytically speaking, there exists a critical
habitat size $L_c$ above which survival of a population is assured.
Here we examine what role the population density plays since, intuitively,
one would expect that even for $L>L_c$, a sufficiently large
population would be needed for growth. 

Lattice based models are widely used in ecology \citep[see for example][]{Tainaka,
Durrett, Itoh} and so we introduce such a model that incorporates birth,
death and diffusion. Unlike other similar models such as the \textit{contact process} \citep[e.g.][]{Harris, Oborny}, here
two individuals must meet in order to reproduce whereas one individual can
die by itself. This results in negative growth rates for populations that
fall below a critical density due to reproduction opportunities becoming
rare. In real populations, the positive correlation between size
and per capita growth rate of a population is known as the Allee effect \citep{Allee},
which has recently received much interest \cite[e.g.][]{Dennis2, Hurford,
Johnson}.
If the Allee effect is strong enough, the population size may even decrease
for small population sizes as in our model. Due to this behaviour, the effect has been examined with respect to extinction
\citep[see for example][and references therein]{Amarasekare, Courchamp, Stephens} but {\em primarily} deterministically and so without fluctuations in the population density.
Since fluctuations are likely to be highly significant for small populations,
we include the effects of these by examining Monte Carlo (MC) simulations
in the hope to gain a more realistic picture of the importance of the population density on the chances of survival. Further to the stochastic methods used to study Allee effects, such as stochastic differential equations \citep[e.g.][]{Dennis},
discrete-time Markov-chains \cite[e.g.][]{Allen} or diffusion processes \citep[e.g.][]{Dennis2}, lattice based models have space, as well as time, as a variable and take into account individuals, rather than just the macroscopic view of the population.
\\ \\
After introducing the model in the next section, we examine the Allee effects
present in our model in Section \ref{Section: Allee Effects}, particularly
with respect to a sudden decrease in population in Section \ref{Section: Decrease in PopDens}. The effects of the fluctuations are examined in Section
\ref{Section: Fluctuations}.

\section{The Model} \label{Section: The Model}
We have a $d$-dimensional square lattice of linear length $L$ where each site is either
occupied by a single particle (1) or is empty (0). A site is chosen at random. If the site is occupied, the particle is removed with probability $p_d$,
leaving the site empty. If the particle does not die, a nearest neighbour site is randomly chosen. If the neighbouring site
is empty, the particle moves to that  site. If however the
neighbouring site is occupied, with probability
$p_b$\footnote{We
note that the birth rate is actually given by $p_b(1-p_d)$ and not $p_b$ only.}, the particle reproduces, producing a new particle on another randomly selected neighbouring
site, conditional on that chosen square being empty. We therefore have the
following reactions for a particle $A$:
        \begin{equation}
        A\phi \longleftrightarrow\phi A,\quad A+A \longrightarrow                3A \quad\mbox{and}\quad A\longrightarrow\phi,
        \end{equation}
where $\phi$ represents an empty site. A time step is defined
as the number of lattice sites and so is equal to approximately one update
per  site. We use nearest neighbours and, throughout most of the paper,
periodic boundary conditions
which, although more unrealistic than, say, reflective boundary conditions, allow for  better comparison with analytical results, since periodic systems
remain homogenous. We later, however, examine some results with reflective boundary conditions.

Due to the conflict between the growth and decay processes in the model, we expect that with certain values of $p_b$ and $p_d$, extinction of the
population would
occur. Indeed, many models displaying such a conflict \citep[see for example ][]{Vespignani, Dammer, Oborny,
Peters} show a critical parameter value separating
an \textit{active} state and an \textit{inactive} or \textit{absorbing}
state that, once reached, the system cannot leave. As the rate of decay
increases, the so-called \textit{order parameter} (often the density of active sites)
decreases, becoming zero at a critical point, marking a change in phase
or \textit{phase transition}. In our case, the absorbing
state would represent an empty lattice and so extinction of the population.

To show that this is indeed the case for our model, we derive a so-called
mean field equation \citep[e.g.][]{Opper} for the density of occupied sites $\rho(t)$. Assuming the particles are spaced homogeneously in an infinite system we have
        \begin{equation} \label{mean field}
        \pderiv{\rho(t)}{t} = p_b(1-p_d)\rho(t)^2(1-\rho(t))-p_d\rho(t).         \end{equation}
The first term is the proliferation term and so is proportional to $\rho^2$,
the probability that the particle does not die $(1-p_d)$, the probability that the next randomly chosen site to give birth on is empty $(1-\rho)$ and
finally the probability that it gives birth if this is the case, $p_b$. The second term represents
particle annihilation and so is proportional to both $\rho$ and $p_d$, the probability
that the chosen particle dies.

Eq. (\ref{mean field}) has three steady states,
        \begin{equation} \label{steady states}
        \bar\rho_0 = 0, \quad \bar\rho_\pm=\frac{1}{2}\left(1\pm\sqrt{1-\frac{4p_d}{p_b(1-p_d)}}\right).
        \end{equation}  
For $4p_d > p_b(1-p_d)$, $\bar\rho_\pm$ are imaginary, resulting in $\bar\rho_0$
being the only real stationary state and so, here, extinction occurs in all
circumstances. Keeping $p_b$ a constant from now on, we then have that our critical death rate is given by $p_{d_c} = p_b/(4+p_b)$ which separates the active
phase representing
survival and the absorbing state of extinction. 

\begin{figure}[tb]
         \centering\noindent
         \begin{tabular}{c}
         \tiny{a)} \\
         \includegraphics[width=8cm]{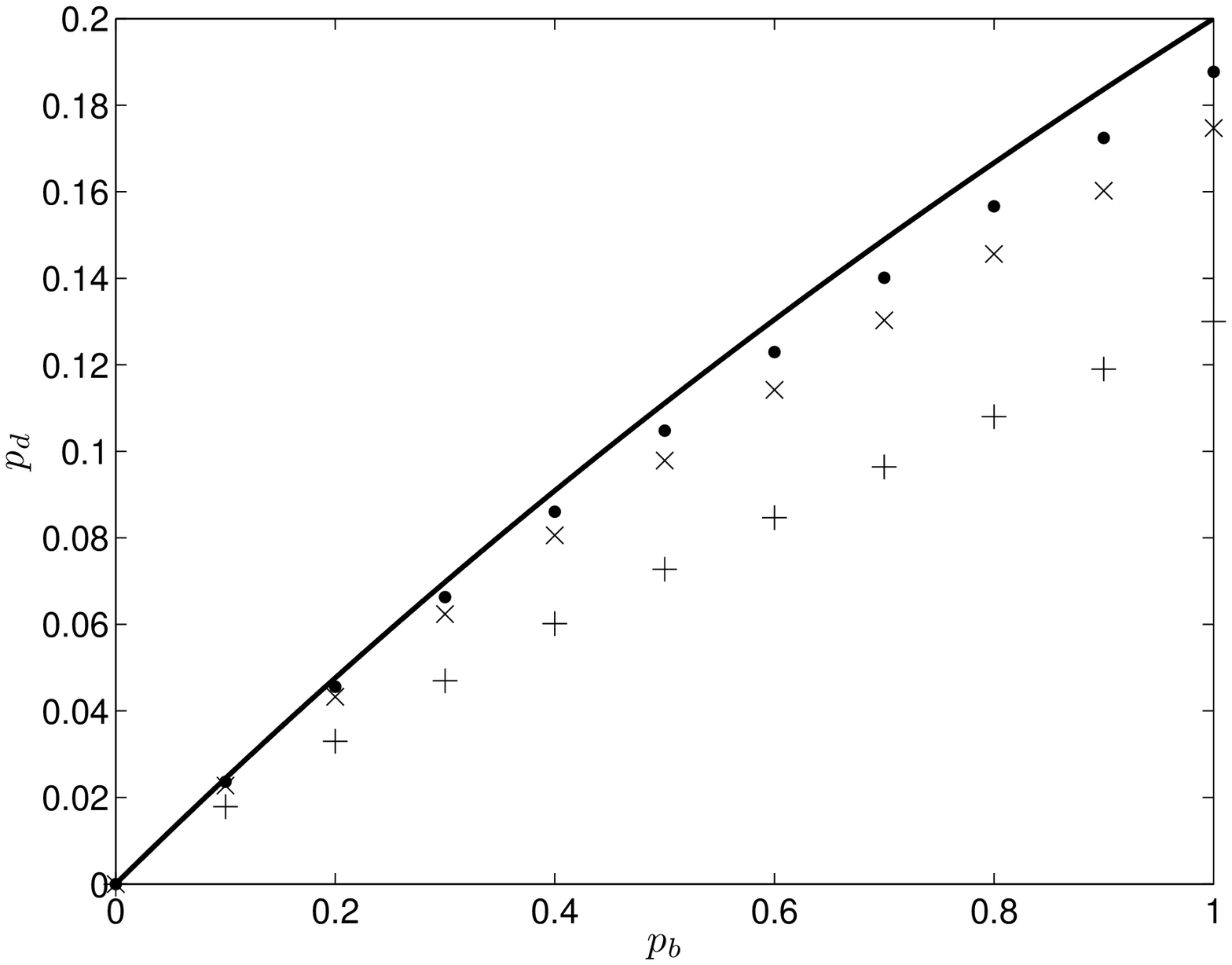} 
         \\ \tiny{b)} \\
         \includegraphics[width=8cm]{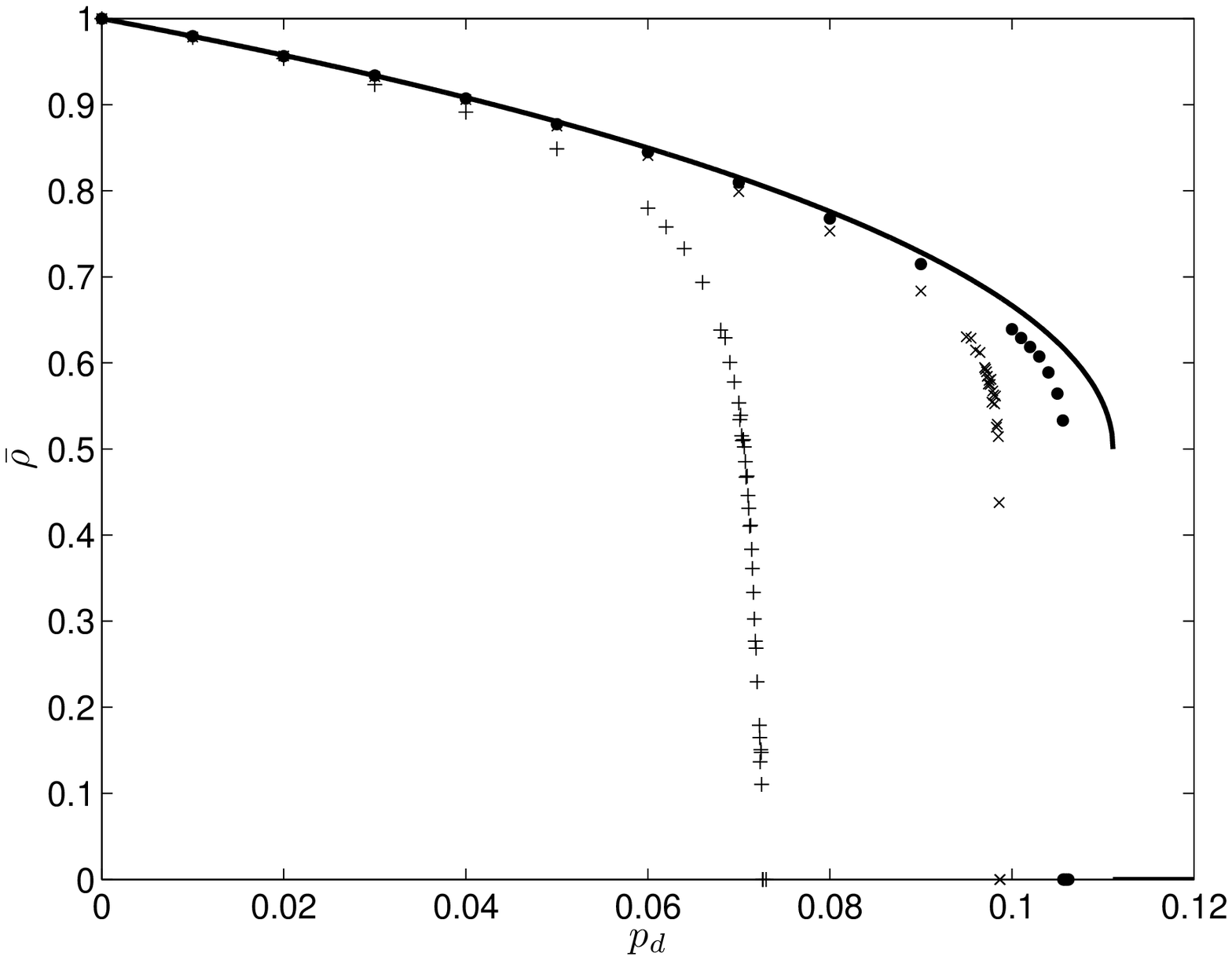} 
         \end{tabular}  
         \caption{a) Parameter domain for the number of steady states with          the points showing the parameters giving 2 steady states  and b)          the steady state population densities with $p_b = 0.5$ for the mean          field (line),
         1 ($+$), 2 ($\times$) and 3 ($\bullet$) dimensional simulations.}
         \label{Critical Parameters}
         \end{figure}

Clearly Eq. (\ref{mean field}) is limited by the exclusion of diffusion and
noise as well as the false assumption of a homogenous population density.  We do
however find at least good qualitative support   for our mean field analysis
through numerical simulations. Fig. \ref{Critical Parameters} a)
shows the critical values of $p_d$ and $p_b$ separating the regions with
one and three stationary states according to both the mean field equation
and numerical simulations for $1$, $2$ and $3$ dimensions.
We see convincing agreement between our analytical and numerical results,
particularly for higher dimensions. The MC simulations were carried out on an initially fully occupied lattice with linear sizes $L=1000$,
$32$ and $10$ for each dimension respectively and we observed whether extinction
occurred during $10^5$ time steps. For each  birth rate, the simulation
was repeated 500 times. If a single run survived, $p_d$ was increased, whereas
if extinction occurred in all runs, $p_d$ was reduced. Using the same initial
seed for the random number generator, an iterative procedure produced a
critical value with accuracy $\pm 2^{-11}$. This iterative procedure
was then repeated 5 times with different seeds and the average taken. Only a small number of repeats
were needed since the largest variance of the values obtained was of the
order of $10^{-8}$. From the figure we find that to 3 d.p. for $p_b=0.5$,
$p_{d_c} = 0.073 $ , 0.098 and 0.105 in 1,2 and 3 dimensions respectively. Due
to the finite size of the lattices and the finite time used for the above simulations, the actual critical death rates are likely to differ slightly
from
those given and more accurate techniques would have to be used to obtain
them \citep[see][for examples of such techniques]{Hinrichsen_Non}.

With $\rho(t = 0) = 1$, as $p_d$ is increased, the steady-state population density decreases, becoming zero at $p_{d_c}$ as shown in Fig. \ref{Critical Parameters}
b), marking the phase transition. We see that the steady
state population density {\em appears} to change continuously in 1 dimension, whilst
discontinuously in 2 and 3 dimensions in agreement with the mean field results.
If indeed this is the case, we call such phase transitions \textit{continuous} and \textit{first-order} respectively. In both cases, the phase transition is marked by a very rapid
decrease in population density.
\\ \\
Briefly relating our model to biology, we note that the death rate for a given species may fluctuate for any number of
reasons but it must certainly
be true that at least the average value of $p_d$ must be less than $p_{d_c}$ for the species to have ever been in existence. However, as we have shown, even a temporary increase in $p_d$ above $p_{d_c}$, due to deforestation
or disease for example, will cause a very rapid and perhaps unrecoverable decrease in population. Extinctions however may also occur for reasons other than having
a super-critical death rate. We investigate the roles
of Allee effects and that of fluctuations in the next three sections where we examine simulations in the sub-critical or active phase and use the constant value \(p_{b}=0.5\).

\section{Allee Effects} \label{Section: Allee Effects}
One reason we observe a decline in population growth at low densities is
due to individuals finding it harder to find a mate. This is empirically known to occur in both plant \citep[e.g.][]{Aizen} and animal \citep[e.g.][]{Lande2} populations. In our model, this aspect is incorporated by the fact that
two individuals are required for reproduction whereas an individual can die
by itself. As density decreases, each
individual therefore finds it increasingly difficult
to find another for reproduction before they die. 
To examine this, we return to our mean field equation (\ref{mean
field}).

\begin{figure}[tb]
        \centering\noindent
        \includegraphics[width=8cm]{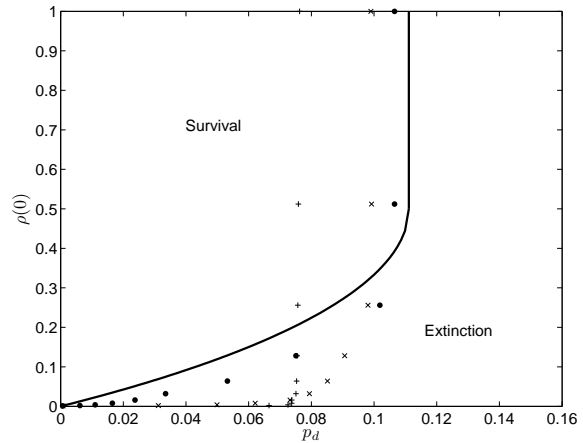}  
        \caption{Phase diagram showing the critical values of  $p_d$
         separating the 2 long-term outcomes of the system for different         initial population density according to the mean field (line) and
        the  1 (+), 2 ($\times$) and 3 ($\bullet$) dimensional MC simulations.}
        \label{phase diagram}
        \end{figure}

It is easy to show that whereas $\bar\rho_+$ and $\bar\rho_0$ are stable stationary points of Eq. (\ref{mean field}), $\bar\rho_-$ is unstable. Since in the active phase, $\bar\rho_0 < \bar\rho_- < \bar\rho_+$, any population whose density $\rho(t)< \bar\rho_-$ will be driven to
extinction by the dynamics of the system. In fact
we find that for $p_d < p_{d_c}$,
        \begin{equation} \label{Long term behaviour}
        \rho(t) \longrightarrow \quad \left\{ 
        \begin{array}{cl}
        0 &  \mbox{ for } \rho(t) < \bar\rho_- \\
        \bar\rho_+ & \mbox{ for } \rho(t) > \bar\rho_-
        \end{array} \right.
        \quad\mbox{ as } t \longrightarrow \infty. 
        \end{equation} 
We test this numerically in 1, 2 and 3 spatial dimensions by finding the
value of $p_d$ that separates the active and absorbing states for different initial conditions. The MC simulations were carried out and the
critical death rate found iteratively in the same fashion as in Section \ref{Section:
The Model}. The results
are shown in Fig. \ref{phase diagram}
and clearly show the importance of the initial population density for survival.
The density dependence appears to increase with 
dimensionality, which we expect, since two individuals meeting becomes progressively
harder as the dimensionality of the system increases.

The existence of this critical population density is highly significant to the conservation of species. It is clear that a sufficiently small population
will not grow, regardless of how much space and resources are available.
It also has repercussions if a population density were to suddenly decrease
due to disease or particularly harsh meteorological conditions, for example. We examine
this further in Section \ref{Section: Decrease in PopDens} after examining
the role of fluctuations.

\section{Fluctuations} \label{Section: Fluctuations}
We expect extinction due to fluctuations in the population density to occur
when the order of the fluctuations approaches the mean population density.
Empirically, demographic stochasticity (that is, chance events of mortality and reproduction) is known to be greater in smaller populations \citep{Lande}
than in larger ones. Population and habitat size are, on average, positively correlated and so, particularly with the existence of the critical population density,
we expect extinction due to fluctuations to occur for smaller lattice sizes as has been suggested
by others \citep[e.g.][]{Pimm, Escudero_Buceta}. 

\begin{figure}[tb]
         \centering\noindent
         \includegraphics[width=8cm]{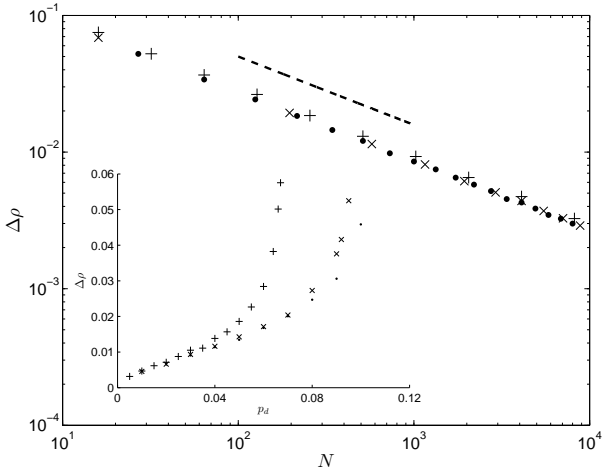} 
         \caption{Log-log plot of the standard deviation of the population
         density versus the number of sites in the 1 (+), 2 ($\times$)
         and 3 ($\bullet$) dimensional systems. The hashed line has gradient
         -0.5 for the eye and indicates the power law behaviour. Insert:          The fluctuations v.s. $p_d$ dimensional case with the same symbol
         notation.}
         \label{Fluctuations}
         \end{figure}

\begin{figure}[tb]
        \centering\noindent
        \includegraphics[width=8cm]{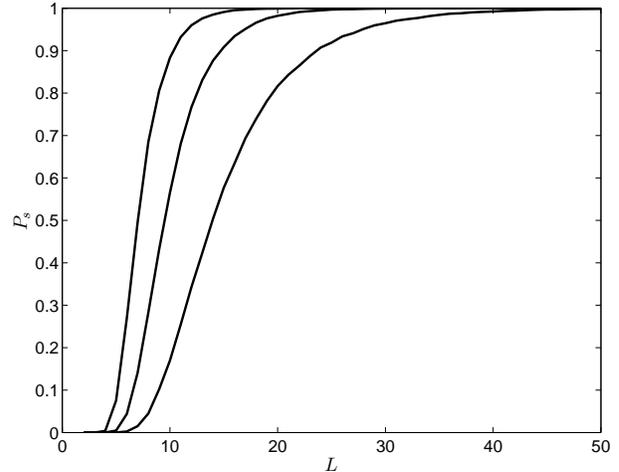}
        \caption{How $P_s$ varies with $L$ for the 1 dimensional model with
        (from left to right), $p_d$ = 0.04, 0.05 and 0.06. Similar results
        are seen in 2 and 3 dimensions.}
        \label{IncreasingL}
        \end{figure}

We see in Fig. \ref{Fluctuations}
that, numerically, the fluctuations in the population density
$\Delta\rho$ decrease with the number of lattice sites $N$ through a power law with exponent -0.50 in all dimensions, which is what we would expect
from the \textit{central limit theorem}. Simulations were carried out for  fixed $p_d=0.03$ and $p_b=0.5$ and the standard deviation obtained from $5\times 10^3$ surviving runs for each lattice size. The insert in Fig. \ref{Fluctuations} shows how the size
of the fluctuations also increase as the critical point is approached. These larger fluctuations will also increase
the probability of extinction as indicated in Fig. \ref{IncreasingL}
where we examine the probability of survival $P_s$, that is, the probability
that extinction has not occurred up to some time $t_m$. We examine the 1
dimensional case only using three different values of $p_d$ with $t_m = 10^3$
and repeat the simulation $5\times10^4$ times for each lattice size. The figures
clearly show how the probability of survival increases with $L$, yet decreases
as $p_d$ increases. Indeed, as $p_{d_c}$ is approached,
population density decreases and fluctuation size increases resulting in
species with higher death rates being more susceptible to extinction.
This is indeed observed in nature where long-lived species are known,
in general, to have a higher chance of survival than short-lived ones \citep{Pimm}.

\section{A Decrease in Population Density} \label{Section: Decrease in PopDens}
Apart from the initial conditions, it is certainly conceivable that the population
density could fall below the critical value due to a reduction in population
size. From Eq. (\ref{Long
term behaviour}) we expect that the population will survive only as long as $\rho(t)>\rho_c$. We simulate this by  increasing  $p_d$ to
1 at some  $t=t_k$ and then returning $p_d$ to what it was before, once a density $\rho_s$
has been reached. We examine this here in 2 dimensions with now reflective
rather than the previously used periodic boundary conditions. Qualitatively,
all previous results have been very similar when using reflective boundary
conditions but here we want to increase this degree of realism in our model.

For 2 dimensional simulations with an initial population density $\rho(0) = 0.128$, the
critical death rate is 0.093 (3 d.p.) as shown in Fig. \ref{phase diagram}. So, for $p_d = 0.093$, we would expect that if $\rho_s
> 0.128$, the population will survive, with the population density returning
to what it was before, whereas for $\rho_s
< 0.128$, extinction will occur. Due to the fluctuations in the population that occur in the simulations, we expect more of an increase in the likelihood
of extinction as $\rho \longrightarrow \rho_c^+$ rather than the definite
survival/extinction result that the mean field predicts.
         \begin{figure}[tb]
         \centering\noindent
         \begin{tabular}{c}
         \tiny{a)} \\ 
         \includegraphics[width=8cm]{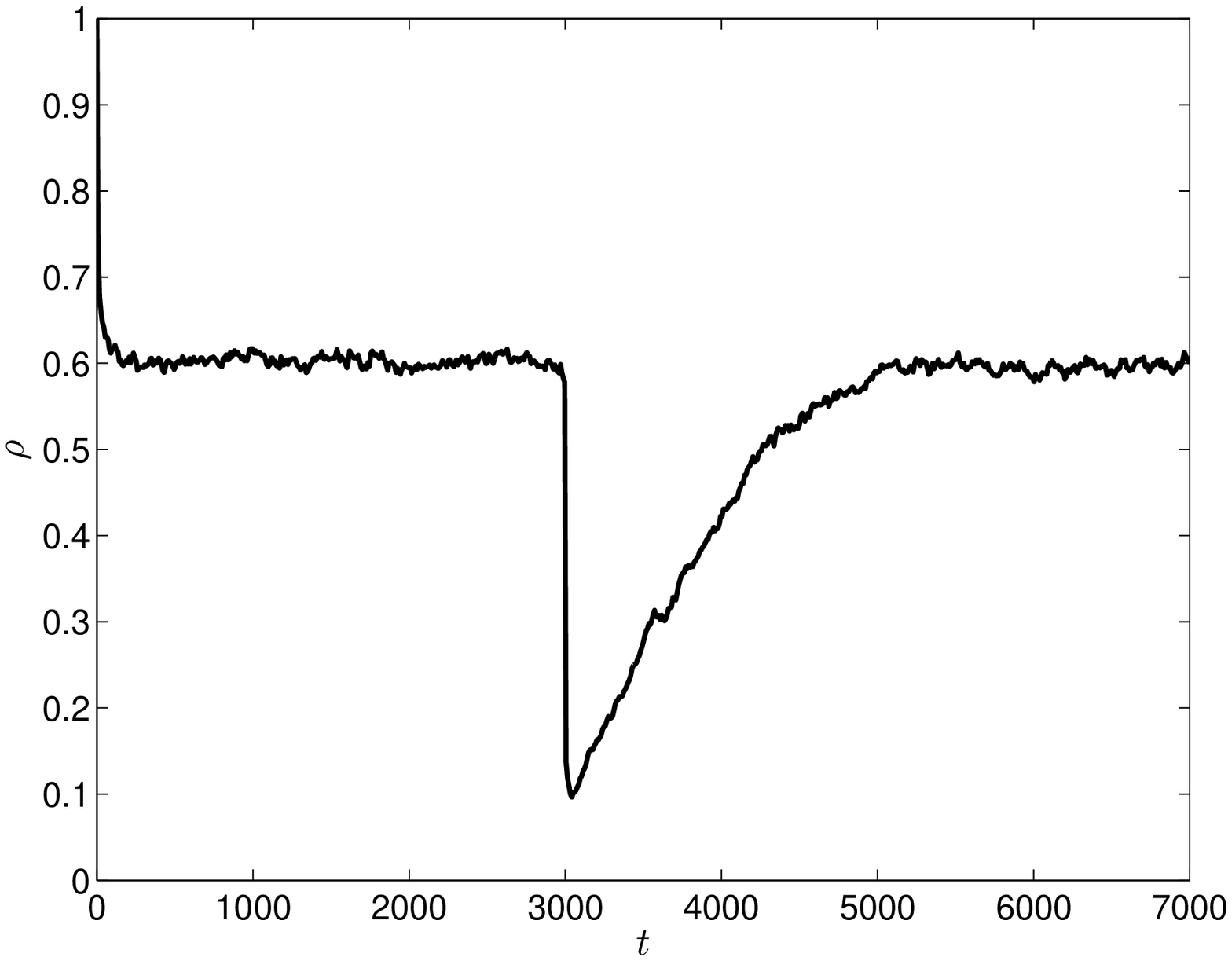}
         \\ \tiny{b)} \\
         \includegraphics[width=8cm]{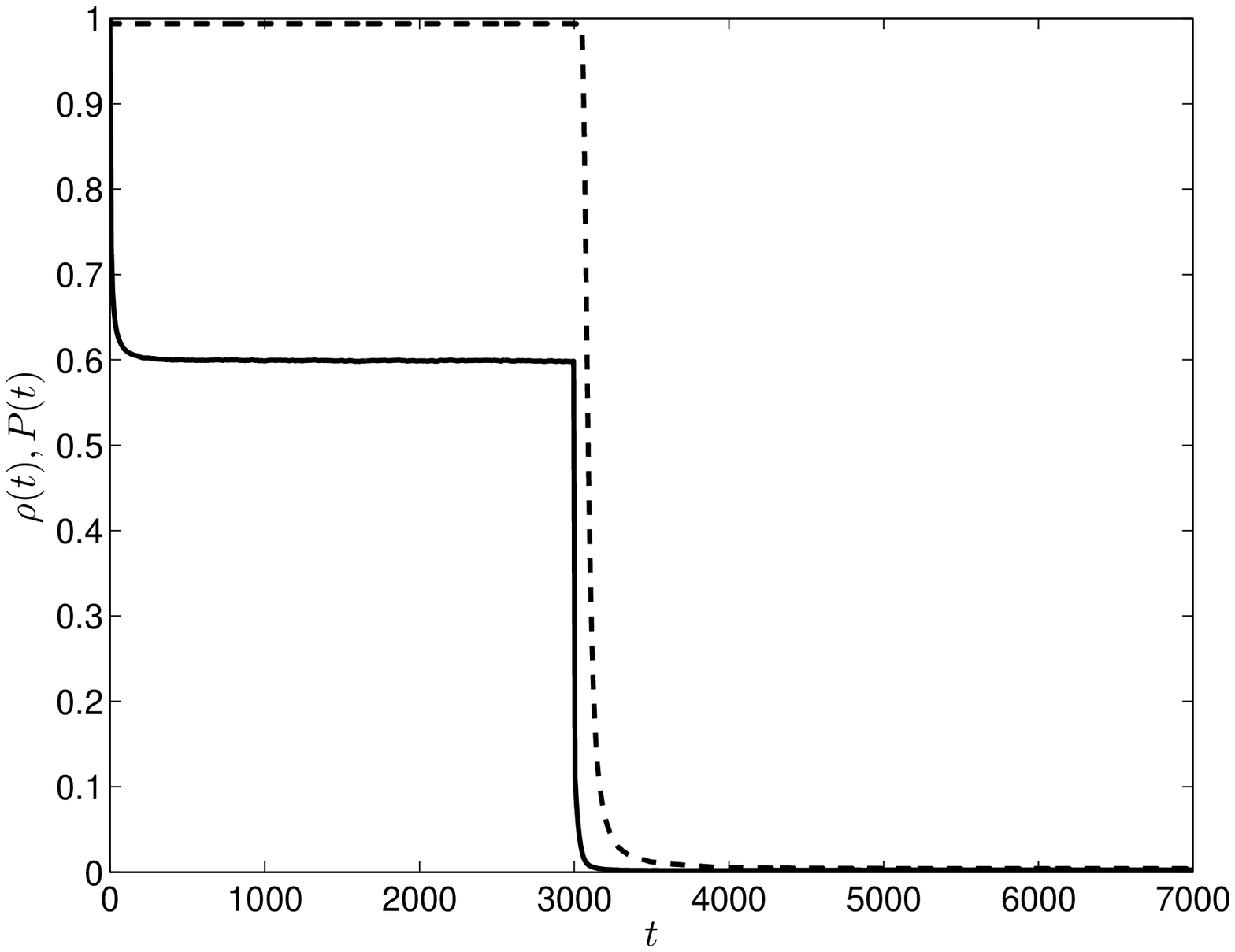} 
         \end{tabular}  
         \caption{a) The average population density of the surviving runs          only.
         b) The average population density of all the runs (solid line) and          the survival probability $P(t)$ (hashed line), i.e. the probability
         that extinction has not occurred up to time $t$.}
         \label{Disease}
         \end{figure}
Fig. \ref{Disease} shows the results for $\rho_s = 0.13$, where we see that for those runs
that {\em did} survive, the population density does indeed return to what it once
was. We also see, as expected, that most of the runs did result in extinction.
In fact the survival rate was 0.004. 

From Fig. \ref{Disease} b) we observe that there is a time delay of approximately
40 time steps between the
sudden decrease in population and when
the survival probability begins to fall. Assuming a particle that dies the $n$th time it is picked, survives
$n-1$ time steps, it is easy to show that the expected lifetime (in time steps) of an individual is given by $(1-p_d)/p_d$. We therefore have a time delay of approximately four lifetimes (recall $p_d = 0.093$) which, for a lot of species, is ample time to act.

In order to prevent extinction in such a case, the population density must be increased beyond $\rho_{c}$. This has important ecological implications
since it shows that the probability of extinction can be decreased, not only
by increasing the population (which is of course not always possible), but
also by a {\em decrease} in habitat size.    

\begin{figure}[tb]
        \centering\noindent
        \includegraphics[width=8cm]{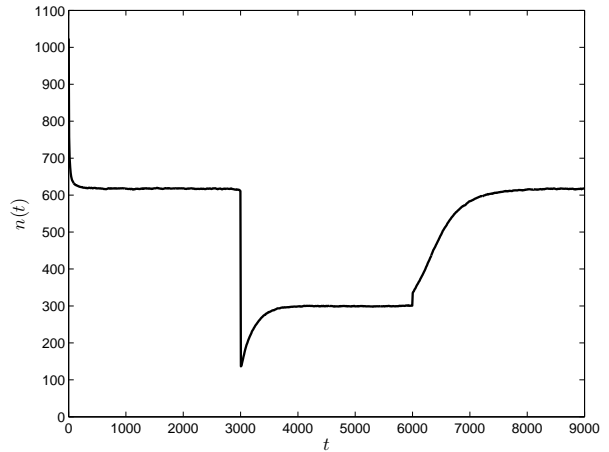}
        \caption{Plot showing the recovery of the population $n(t)$ for
        the surviving runs only after         a disease
        breakout at $t = 3000$ due to the re-sizing of the lattice. The lattice
        is returned to how it was originally at $t=6000$ and the population         recovers its original value.}
        \label{Disease Recovery}
        \end{figure}

To see whether this hypothesis holds, we simulate this again using $p_d
= 0.093$ but this time $\rho_s =\rho_c = 0.128$ so that the chance of survival is negligible. This time however, once the population density has been reduced,
the area covered by the lattice is reduced by half. The organisms in the half that remains are left where they are, whereas those
in the half that is removed are randomly placed in the remaining half. This
then doubles the population density, bringing the population out
of the sub-critical population density. Once the population has recovered
and stabilised, the lattice size is returned to how it once was. The results
are shown in Fig. \ref{Disease Recovery}
and clearly show the recovery of the population once the lattice
size has been reduced. In fact, out of 1000 runs, the probability of survival
rose from 0.003 to 0.281.

\begin{figure}[tb]
        \centering\noindent
        \includegraphics[width=8cm]{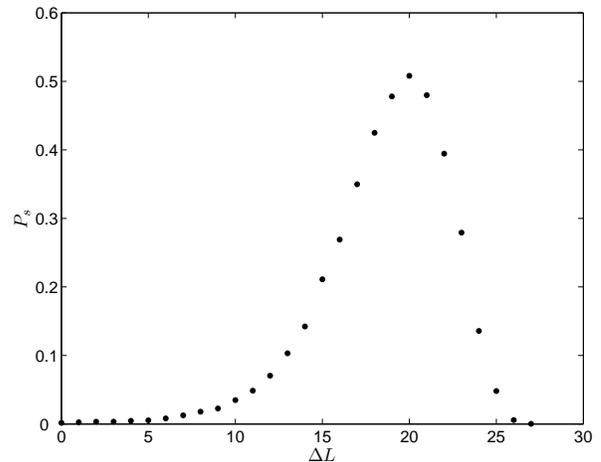}
        \caption{How the probability of survival changes with different reductions
        in $L$, starting from $L=32$.}
        \label{HabitatReduction}
        \end{figure}

We expect there to be an optimal habitat reduction size - too large a reduction
and the population will be in danger from large fluctuations associated with
smaller habitat sizes whereas too small
a reduction and the density will not be increased sufficiently. We therefore
plot in Fig. \ref{HabitatReduction}
the probability that the system does not go extinct up to some $t=t_{max}$,
$P_s$, against the reduction in $L$, $\Delta L$. We again use $p_d = 0.093$
and $\rho_s = 0.128$. For small $\Delta L$, $P_s$ changes little
due to the density not being reduced enough, yet for larger $\Delta L$, the
larger fluctuations resulting from the smaller value of $L$ also cause $P_s$
to be small. 

Whilst reflective boundary conditions were used
here, very similar results were obtained using  periodic boundary conditions.
In fact, with periodic boundary conditions, the probability of survival increased  more significantly by
the decrease in $L$ due to the population being able to grow in two directions
rather than in just one after the habitat size has been returned to what
it once was. This of course could be achieved in reality
by reducing the habitat from more than one direction.
\\ \\
This model was proposed to represent how the area in
which a population is found could be reduced in real-life. The species could
be driven towards one end of the habitat with a boundary placed to prevent
them leaving the desired area. This boundary could then be removed once the
population has recovered. Clearly this is easier for larger, land-based animals
but in principle, at least, could be achieved for all species. 

\section{Conclusions} \label{Section: Conclusions}
Allee effects are certainly observed in nature \cite[][et al]{Stephens2,Pederson,Gyllenberg}
and have been studied with respect to extinction. Using a lattice model,
we have observed Allee effects together with the role of fluctuations, with the advantage of being able to examine the effects of habitat size. Being
able to model the population as a group of \textit{individuals} which move, breed and die, rather than as a variable in an equation, has enabled us to gain a more realistic insight into how real populations behave. 

Rather than the clear-cut conclusions that deterministic models produce,
conservationists often examine the \textit{probability}
that a population will maintain itself without significant demographic or
genetic manipulation for the foreseeable ecological future \citep{Soule}.
In this spirit, for a sufficiently large population density we have shown that the probability of survival does increase with habitat
size due to the smaller fluctuations. However, far more important are the death rate and population density since if these fall on the wrong side of their critical values, extinction is almost a certainty.

Our findings are certainly significant for the design of habitats. The notion of a critical habitat size, mentioned in the Introduction, is misleading, since, it is certainly not true that for a fixed population size, the larger the habitat size the better. Regardless of the amount of space and resources available,
a population will only grow if the density is above its critical
value. We also proposed, in the last section, a method for greatly reducing the probability of extinction by reducing the habitat size once a species has become rare.

\begin{figure}[tb]
        \centering\noindent
        \includegraphics[width=8cm]{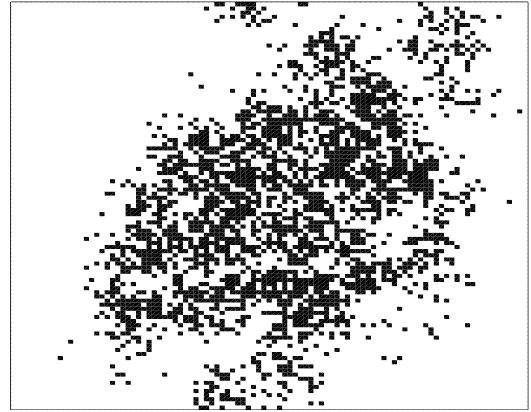}
        \caption{Snapshot of the output from a 2 dimensional lattice with
        $L=100$. A value of $p_d = 0.1$ was used and the picture was taken
        at $t = 600$ when $\rho = 0.2223.$}
        \label{ClusteringPicture}
        \end{figure}

Our notion of density has been that of the number of individuals per unit
area. While we assumed this to be constant in space when deriving our mean
field equation (\ref{mean field}), clearly this will vary amongst real populations. In fact, for populations that are found in patches, the value
of the density will depend very much on the scales used. The same is true
of the MC results as shown in Figure \ref{ClusteringPicture},
where we see clear examples of clustering. In nature, species will cluster
to varying degrees and hence the value of the critical population density
will also vary and would need to be estimated in each case.
\\ \\
Compared to other stochastic models, we claim the use of lattice models gives
a more realistic insight into the way in which real populations behave. We
do still however, recognise the inaccuracies in our model and the difficulties in
implementing the observations. We believe the model to be valid, to a greater
or lesser degree to all species which rely on others for growth, perhaps
particularly those who live alone yet sexually reproduce. In fact, due to the great
variety of species, we have presented the above as ideas which may be of qualitative, rather than quantitative, relevance to conservation management.     
\section*{Acknowledgments}
We would like to thank Be\'ata Oborny for very helpful discussions and references. Alastair Windus would also like to thank the Engineering
and Physical Sciences Research Council (EPSRC) for his Ph.D. studentship. 

\end{document}